\documentclass[prd,reprint,amsmath,amssymb,preprintnumbers,superscriptaddress,nofootinbib]{revtex4-1}
\usepackage{graphicx}  % needed for figures
\usepackage{dcolumn}   % needed for some tables
\usepackage{bm}        % for math
\usepackage{amssymb}   % for math
\usepackage{amsmath}
\usepackage{subcaption}
\newcommand{\eqal}[1]{\begin{align}#1\end{align}}
\def\eq{\begin{equation}}
\def\eqe{\end{equation}}
\def\eqa{\begin{eqnarray}}
\def\eqae{\end{eqnarray}}
\def\p{\partial}

\begin{document}

% The following information is for internal review, please remove them for submission
%\widetext
%\leftline{Version xx as of \today}
%\leftline{Primary authors: Joe E. Physics}
%\leftline{To be submitted to (PRL, PRD-RC, PRD, PLB; choose one.)}
%\leftline{Comment to {\tt d0-run2eb-nnn@fnal.gov} by xxx, yyy}
%\centerline{\em D\O\ INTERNAL DOCUMENT -- NOT FOR PUBLIC DISTRIBUTION}

% the following line is for submission, including submission to the arXiv!!
%\hspace{5.2in} \mbox{Fermilab-Pub-04/xxx-E}

\title{Infinite Soft Theorems from Gauge Symmetry}
      % D0 authors (remove the first 3 lines
                             % of this file prior to submission, they
                             % contain a time stamp for the authorlist)
                             % (includes institutions and visitors)
\date{\today}

\author{Zhi-Zhong Li}
\email{b02202003@ntu.edu.tw}
\affiliation{
Department of Physics, National Taiwan University
}
\author{Hung-Hwa Lin}\email{hunghwalin@gmail.com}
\affiliation{
Department of Physics, National Taiwan University
}
\author{Shun-Qing Zhang}\email{e24019025@gmail.com}
\affiliation{
Department of Physics, National Taiwan University
}

\begin{abstract}
In this letter we show that the soft behaviour of photons and graviton amplitudes, after projection, can be determined to infinite order in soft expansion via ordinary on-shell gauge invariance. In particular, as one of the particle's momenta becomes soft, gauge invariance relates the non-singular diagrams of an $n$-point amplitude to that of the singular ones up to possible homogeneous terms. We demonstrate that with a particular projection of the soft-limit, the homogeneous terms do not contribute, and one arrives at an infinite soft theorem. This reproduces the result recently derived from the Ward identity of large gauge transformations. We also discuss the modification of these soft theorems due to the presence of higher-dimensional operators.  
\end{abstract}

\maketitle

\section{Introduction}
It has long been known that on-shell gauge invariance can be utilized to obtain universal soft behaviours of scattering amplitudes for photons and gravitons. Gauge invariance dictates that the amplitude must vanish when one of its polarization vector/tensor is replaced by the momenta. Taking one of the momenta of an $n$-point amplitude ($M_{n}$) to be soft, the gauge invariance of the soft leg then relates the finite part of the amplitude to the singular diagrams, which is given by the product of a three-point vertex and the $n{-}1$-point amplitude. The latter is then amenable to the form of a ``soft operator" acting upon the $n{-}1$-point amplitude. Thus one schematically have:
\eq
M_{n}|_{q\rightarrow 0}=\sum_{i=-1}^a\left(\mathcal{S}_{i}\right)M_{n{-}1}+\mathcal{O}(q^{a+1})
\eqe
where $q$ is the soft momenta, and $\mathcal{S}_{i}$ are the soft operators with its subscript indicating to which degree in the $q$ expansion is it defined. For photons $a=0$, while for gravitons $a=1$~\cite{LowFourPt, LowTheorem, Weinberg, OtherSoftPhotons}. The reason why the soft theorem always terminate at a finite order is because when using gauge invariance, one can only determine the finite part of the amplitude up to a homogeneous solution, denoted as $R^\mu$, satisfying $q\cdot R=0$ for which one has no control. From general principle of locality and Lorentz symmetry, one can only determine the minimum order in $q$ must this term contain, which sets $a$. 

A few years ago, Strominger and
collaborators~\cite{Strominger,CachazoStrominger} demonstrated that the soft-theorems for gravitons can alternatively be interpreted as a consequence of  extended Bondi, van der Burg, Metzner and Sachs (BMS) symmetry~\cite{BMS,
  ExtendedBMS}. This generated new interest in soft-theorems of amplitudes and its relationship with underlying symmetry. As the new interpretation only relies on the structure of space-time at asymptotic infinity, it can be viewed as a direct constraint on any theory of quantum gravity which admits asymptotic flat solutions. However, given that the resulting soft theorems can be derived via ordinary gauge symmetry, it is natural to ask, in the context of amplitudes, what precisely does the new interpretation buy us? This is especially intruiguing given that the soft theorems are modified at loop-level~\cite{BernLoop, Yutin} as well as higher-dimensional operators~\cite{Yutin2, elvang}, which are tied to the details of the interaction.

Recently, an interesting opportunity presented itself in the form of an infinite order soft-theorem derived from the Ward identity of large gauge transformations \footnote{See also \cite{Schwab} for the derivation of a Ward identity for residual gauge symmetry.} by Hamada and Shiu~\cite{gary}. An interesting feature of the newly derived soft-theorem is that it only gives the soft-limit of the projected piece of the amplitude. For example for photons one has:
\eq
\left.\Omega_{\mu\alpha_{1}\cdots\alpha_{l}} \partial_{q^{\alpha_{1}}}\cdots\partial_{q^{\alpha_{l}}} M_{n}^{\mu}\right |_{q\rightarrow 0}=\sum_{i=-1}^\infty\left(\mathcal{S}^{\mu}_{i,\nu}\right)M^\nu_{n{-}1}
\eqe
where $M_{n}^{\mu}$ is the amplitude with one of the polarisation vector $\epsilon^\mu$ stripped off and $\Omega_{\mu\alpha_{1}\cdots\alpha_{l}}$ is a symmetric tensor.\footnote{\label{note:traceless} In~\cite{gary}, the tensor is symmetric traceless. However the trace piece automatically vanishes upon contracting with the polarisation vectors, as we will discuss shortly, and thus do not make a difference.} 

In this letter, we will show that the above can again be derived by ordinary on-shell gauge invariance. Recall that the derivation based on gauge symmetry yields soft theorems at finite order due to the potential ambiguities, i.e. the aforementioned $R^\mu$. We will demonstrate that such terms vanish upon the projection. In other words, the infinite order soft-theorem derived in~\cite{gary} is precisely the part of the amplitude that are completely determined by ordinary gauge symmetry. We will demonstrate this for photon and gravitons. Furthermore, we will use explicit examples to demonstrate that while $R^\mu$ can be projected out, it is nonetheless not zero. Finally, for completeness we will discuss the modification of this infinite soft theorem by the presence of higher dimensional operators.

\section{Soft Theorem from Ward Identity}
We follow \cite{bern,plefka} to investigate infinite order soft limits of photon and graviton amplitude using ordinary on-shell gauge invariance. 
Beyond the usual (sub)subleading soft theorems, they could only be fixed up to a homogeneous term. 
However, if restrict our attention certain projected pieces of the amplitude, such term does not contribute, and soft theorems can be obtained up to infinite order. 
For photons, we reproduce the result from large gauge transformations \cite{gary}. 
For gravitons, our result is more general, in that it gives the soft limit of a broader piece of the amplitude. That is, the soft theorems here left fewer undetermined pieces than the result in \cite{gary}. 

\subsection{Photon Soft Theorem}

Consider a scattering amplitude 
\eqal{
M_{n+m+1}\left(q;p_{1},\cdots,p_{m},k_{1},\cdots k_{n}\right)
}
involving one soft photon, $n$ hard photons, and $m$ matter
scalars, with momenta $q$, $k_{1},\cdots k_{n}$, and $p_{1},\cdots,p_{m}$,
respectively. Since the amplitude is a linear function in polarization
vectors, it can be expressed as
\begin{align}
M_{n+m+1}= & \epsilon_{q,\mu}M_{n+m+1}^{\mu} \,,
\label{eq:Amp_eps}
\end{align}
where $\epsilon_q$ is the polarization vector for the soft photon.
In the following we discuss the partial amplitude $M_{n+m+1}^{\mu}$
without the polarization vector.

\begin{figure}[h]
\begin{subfigure}[b]{0.4\linewidth}
\includegraphics[height=0.6\textwidth]{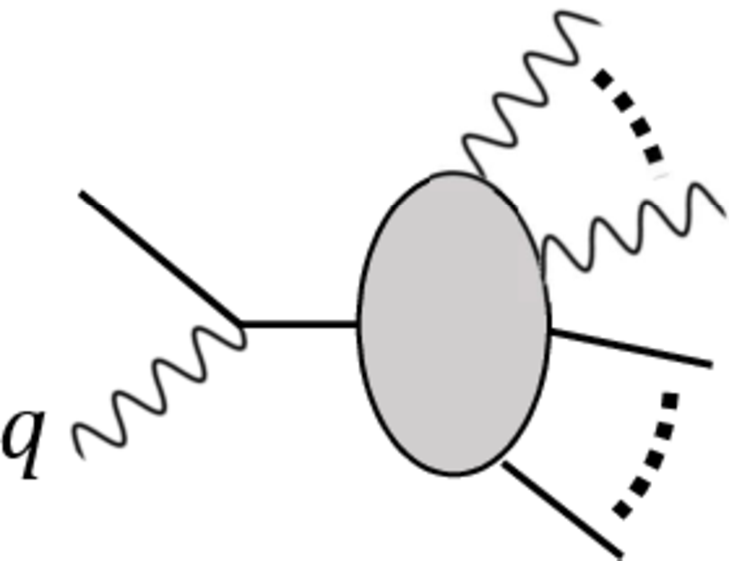}
\caption{} 
\label{fig:photons_pole}
\end{subfigure}
%%%
\begin{subfigure}[b]{0.4\linewidth}
\includegraphics[height=0.62\textwidth]{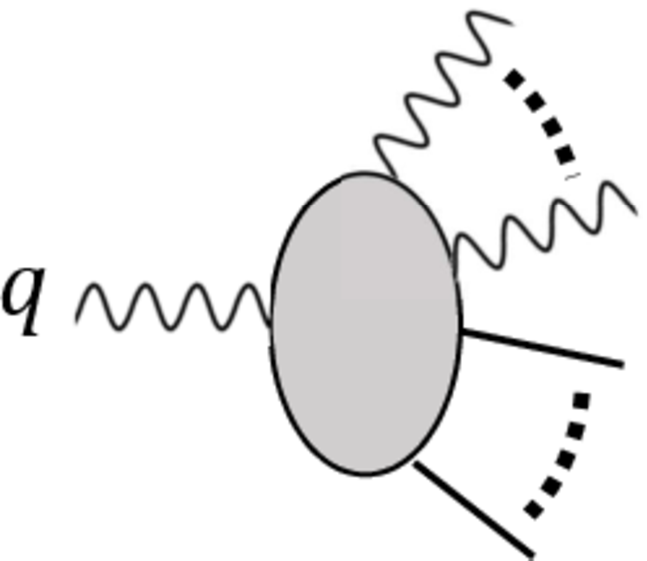}  
\caption{}
\label{fig:photons_gut}
\end{subfigure}
\caption{Contributions (a) with pole, (b) without pole. }
\label{fig:photons}
\end{figure}

The scattering amplitude contains contribution with a pole in the
soft momentum $q$ and those with no pole, as in Fig.\ref{fig:photons}, 
\begin{align}
 & M_{n+m+1}^{\mu}\left(q;p_{1},\cdots,p_{m},k_{1},\cdots k_{n}\right)\nonumber \\
= & \sum_{i=1}^{m}e_{i}\frac{p_{i}^{\mu}}{p_{i}\cdot q}M_{n+m}\left(p_{1},\cdots,p_{i}+q,\cdots,p_{m},k_{1},\cdots k_{n}\right)\nonumber \\
 & +N^{\mu}\left(q;p_{1},\cdots,p_{m},k_{1},\cdots k_{n}\right)\,,
 \label{eq:M_pole}
\end{align}
where $e_{i}$ are the charges of scalars, $N^{\mu}$ denotes the terms
without pole, and $M_{n+m}$ denotes the lower point amplitude without
the soft photon. The pole terms can only arise from the three point vertex involving
the soft photon and an external scalar, since there are no self interaction
for photons.
At leading order, there is no contribution from $N^{\mu}$, giving the leading soft theorem 
\eqal{
M_{n+m+1}^{\mu}\big|_{q\rightarrow 0} =&\sum_{i=1}^{m}e_{i}\frac{p_{i}^{\mu}}{p_{i}\cdot q}M_{n+m}+\mathcal{O}\left(q^{0}\right) \,.
}
Beyond this order, $N^{\mu}$ must be considered. 

On-shell gauge invariance relates $N^{\mu}$ to the lower-point amplitude $M_{n+m}$ by dictating
\begin{align}
0= & q_{\mu}M_{n+m+1}^{\mu}=\sum_{i=1}^{m}e_{i}M_{n+m}+q_{\mu}N^{\mu}.
\label{eq:Ward_N}
\end{align} 
At zeroth order, the constraint gives charge conservation, 
\begin{align}
\sum_{i=1}^{m}e_{i}= & 0 \, .
\label{eq:photon_leading}
\end{align}
Beyond zeroth order, we may expand $N^{\mu}$ as 
\begin{align}
N^{\mu}= & \sum_{l}q_{\alpha_{1}}\cdots q_{\alpha_{l}}N_{l}^{\mu,\alpha_{1}\cdots\alpha_{l}}\,,
\label{eq:N_expand}
\end{align}
since it is polynomial in $q$ at tree level. 
Then, order by order we have 
\begin{align}
q_{\mu} & q_{\alpha_{1}}\cdots q_{\alpha_{l}} \times \nonumber \\
 & \left( \sum_{i=1}^{m} \frac{e_{i}}{(l+1)!}\partial{}_{i}^{\mu}\partial_{i}^{\alpha_{1}}\cdots\partial_{i}^{\alpha_{l}}M_{n}+N_{l}^{\mu,\alpha_{1}\cdots\alpha_{l}}\right)=0 \,.
\end{align}
so that $N_{l}^{\mu}$ can be expressed in terms of $M_{n+m}$ up to a homogeneous term $R_{l}$,
\begin{align}
N_{l}^{\mu,\alpha_{1}\cdots\alpha_{l}}= & -\sum_{i=1}^{m} \frac{e_{i}}{(l+1)!}\partial_{i}^{\mu}\partial_{i}^{\alpha_{1}}\cdots\partial_{i}^{\alpha_{l}}M_{n+m}\nonumber \\
 & +R_{l}^{\mu\alpha_{1}\cdots\alpha_{l}} \,,
 \label{eq:Ward_N_l}
\end{align}
where $R_{l}$ satisfies Ward identity by itself 
\begin{align}
q_{\mu}q_{\alpha_{1}}\cdots q_{\alpha_{l}}R_{l}^{\mu\alpha_{1}\cdots\alpha_{l}} & =0 \,,
\label{eq:Ward_homo}
\end{align}
posing as an ambiguous term. Generally, $R_{l}$ can be separated into three pieces, 
\eqal{
R^{\mu\alpha_{1}\cdots\alpha_{l}} &= T_{l}^{\mu\alpha_{1}\cdots\alpha_{l}}+O_{l}^{\mu\alpha_{1}\cdots\alpha_{l}}+A_{l}^{\mu\alpha_{1}\cdots\alpha_{l}} \,,
}
where $T$ is the trace part, 
\eqal{
T_{l}^{\mu\alpha_{1}\cdots\alpha_{l}} &= \eta^{(\mu\alpha_{1}}B_{l}^{\alpha_{2}\cdots\alpha_{l})} \,,
} 
$O$ is the symmetric traceless part satisfying 
\eqal{
\eta_{\mu\alpha_{i}}O_{l}^{\mu\alpha_{1}\cdots\alpha_{l}}=\eta_{\alpha_{i}\alpha_{j}}O_{l}^{\mu\alpha_{1}\cdots\alpha_{l}}=0,\text{ for any }i,j \,,
}
and $A$ contains the remaining terms, which are antisymmetric in any two indices among $\mu$ and $\alpha$'s. 
Since any arbitrary $A$ or $T$ automatically satisfy Eq.\,\eqref{eq:Ward_homo},
the symmetric traceless part $O$ must satisfy Eq.\,\eqref{eq:Ward_homo} by itself.
It is then straightforward to show that $O$ must vanish \footnote{For arbitrary $p$, we have following separations: $p_\mu= \sum_{i=1}^3 c_i q_{i\mu}$ and $p_\mu p_\nu = c_0 \eta_{\mu \nu} + \sum_{i=1}^3 c_i q_{i\mu} q_{i\nu}$, where $q_i^2 = 0$.}. 
The trace part $T$ can also be discarded, since the contribution of $R_{l}$ to $N^{\mu}$ is in the form of
\begin{align}
q_{\alpha_{1}}\cdots q_{\alpha_{l}}R_{l}^{\mu\alpha_{1}\cdots\alpha_{l}} \,,
\end{align}
so that $T$ either produces terms with $q^{2}=0$ for massless $q$, or $q^{\mu}$ which vanishes after putting back the polarization vector of the soft photon, as in Eq.\,\eqref{eq:Amp_eps}. 
Therefore, only the antisymmetric part need to be considered, giving us 
\begin{align}
N_{l}^{\mu,\alpha_{1}\cdots\alpha_{l}}= & -\sum_{i=1}^{m} \frac{e_{i}}{(l+1)!}\partial_{i}^{\mu}\partial_{i}^{\alpha_{1}}\cdots\partial_{i}^{\alpha_{l}}M_{n+m}\nonumber \\
 & +A_{l}^{\mu\alpha_{1}\cdots\alpha_{l}} \,.
 \label{eq:Ward_N_l-1}
\end{align}
Plugging this into the expression for full amplitude Eq.\,\eqref{eq:M_pole},
we get an incomplete soft theorem for all orders up to the antisymmetric
homogeneous term $A$, 
\eqal{
&M_{n+m+1,\left(l\right)}^{\mu}\nonumber\\
=&\sum_{i=1}^{m}\frac{1}{(l+1)!}\frac{e_{i}}{p_{i}\cdot q}q_{\nu}J_{i}^{\mu\nu}\left(q\cdot\partial_{i}\right)^{l}M_{n+m}\nonumber\\
&+q_{\alpha_{1}}\cdots q_{\alpha_{l}}A_{l}^{\mu\alpha_{1}\cdots\alpha_{l}} \,.
}
where 
\begin{align}
J_{i}^{\mu\nu}= & p_{i}^{\mu}\frac{\partial}{\partial p_{i\nu}}-p_{i}^{\nu}\frac{\partial}{\partial p_{i\mu}},
\end{align}
The case $l=0$ contains no homogeneous term, giving us the well-known subleading soft theorem. 
At higher order $A$ can be non-zero, but we may single out the piece totally symmetric in $\alpha_{i}$ and
$\mu$ by contracting with a totally symmetric tensor $\Omega_{\mu\alpha_{1}\cdots\alpha_{l}}$.
$A$ is then removed, giving a partial soft term up to all order in
$q$, 
\begin{align}
&\Omega_{\mu\alpha_{1}\cdots\alpha_{l}} \partial^{\alpha_{1}}\cdots\partial^{\alpha_{l}} M_{n+m+1}^{\mu} \bigg|_{q\rightarrow 0} \nonumber\\
=&\Omega_{\mu\alpha_{1}\cdots\alpha_{l}}\partial^{\alpha_{1}}\cdots\partial^{\alpha_{l}}\nonumber\\&\ \left[\sum_{i=1}^{m}\frac{1}{(l+1)!}\frac{e_{i}}{p_{i}\cdot q}q_{\nu}J_{i}^{\mu\nu}\left(q\cdot\partial_{i}\right)^{l}M_{n+m}\right] \bigg|_{q\rightarrow 0}
\end{align}
where we adopt short-hand notation $\partial^{\alpha_{j}}=\partial/\partial q^{\alpha_{j}}$ and $q\cdot\partial_{i}=q\cdot\partial/\partial p_{i}$. 
These are exactly the infinite order soft theorems in \cite{gary}. 

\subsection{Graviton}

The derivation for soft theorems of gravitons is similar, except that Ward identity can be applied twice, 
pushing the usual soft theorem to subsubleading order, and placing more stringent constraint on the homogeneous terms at higher order. 

In principle, we should consider a general amplitude involving one soft graviton, $n$ hard gravitons, and $m$ matter
scalars, 
\eqal{
M_{n+m+1} \left(q,p_1,\cdots ,p_m, k_1, \cdots, k_n \right),
}
with momenta $q$, $k_{1},\cdots k_{n}$, and $p_{1},\cdots,p_{m}$,
respectively. 
The pole contribution could then come from both the scalar-graviton vertex and the three-point self-interaction of gravitons. 
Though the derivation procedure is unchanged, this complicates the calculation of soft factors. 
For clarity, we separately consider two cases: one involving only a single graviton, and one involving multiple gravitons without scalars. 
The most general soft theorem can be obtained simply by combining the result of the two. 

We first discuss the amplitude involving a single soft graviton and $m$ scalars, with momenta $q$ and $p_1, \cdots, p_m$, respectively, 
\eqal{
M_{m+1} \left( q, p_1, \cdots, p_m \right)
}
The scattering amplitude again contains contribution with and without a pole in the soft momentum $q$,
\begin{align}
 & M_{m+1}^{\mu\nu}=\sum_{i=1}^{m}\frac{p_{i}^{\mu}p_{i}^{\nu}}{p_{i}\cdot q}M_{m}+N^{\mu\nu}
 \label{eq:1grav_polegut}
\end{align}
with $N^{\mu \nu}$ denoting the terms without pole and $M_{n}$ the lower point amplitude without the soft graviton. 
Expanding in the power of soft momentum $q$, only the pole diagrams contribute to the leading piece 
\begin{align}
M_{(-1)}^{\mu\nu}=\sum_{i=1}^{m} \frac{p_{i}^{\mu}p_{i}^{\nu}}{p_{i}\cdot q} \,.
\end{align}
However, the higher order pieces contain both pole and gut diagrams,
and, by Ward identity, parts of gut diagrams relate to the pole ones.
\begin{align}
q_{\mu}\left( \sum_{i=1}^{m} \frac{p_{i}^{\mu}p_{i}^{\nu}}{p_{i}\cdot q}M_{m}(p_{i}+q)+N^{\mu\nu}\right)=0 \,.
\end{align}
Expanding $N^{\mu\nu}$ around $q\rightarrow0$, 
\begin{align}
N^{\mu\nu}=\sum_{l}q^{\alpha_{1}}\cdots q^{\alpha_{l}}N_{l}^{\mu\nu,\alpha_{1}\cdots\alpha_{l}} \,,
\end{align}
we similarly obtain $N_{l}$ up to a homogeneous term $R_{l}$, 
\begin{align}
N_{l}^{\mu\nu,\alpha_{1}\cdots\alpha_{l}}= & -\sum_{i=1}^{m} \frac{p_{i}^{\nu}}{(l+1)!}\partial_{i}^{\mu}\partial_{i}^{\alpha_{1}}\cdots\partial_{i}^{\alpha_{l}}M_{m}\nonumber \\
 & +R_{l}^{\mu\nu\alpha_{1}\cdots\alpha_{l}} \,,
\end{align}
where 
\begin{align}
q_{\mu}q_{\alpha_{1}}\cdots q_{\alpha_{l}}R_{l}^{\mu\nu\alpha_{1}\cdots\alpha_{l}} & =0 \,.
\end{align}
Again, $R_{l}$ can be separated into three pieces, 
\begin{align}
R^{\mu\nu\alpha_{1}\cdots\alpha_{l}} & =T_{l}^{\underline{\mu}\nu\underline{\alpha_{1}\cdots\alpha_{l}}}+O_{l}^{\underline{\mu}\nu\underline{\alpha_{1}\cdots\alpha_{l}}}+A_{l}^{\underline{\mu}\nu\underline{\alpha_{1}\cdots\alpha_{l}}} \,,
\end{align}
where $T$ is the trace part, 
\begin{align}
T_{l}^{\underline{\mu}\nu\underline{\alpha_{1}\cdots\alpha_{l}}} & =\eta^{(\mu\alpha_{1}}B_{l}^{\alpha_{2}\cdots\alpha_{l})\nu} \,,
\end{align}
$O$ is the symmetric traceless part satisfying
\begin{align}
\eta_{\mu\alpha_{i}}O_{l}^{\underline{\mu}\nu\underline{\alpha_{1}\cdots\alpha_{l}}}=\eta_{\alpha_{i}\alpha_{j}}O_{l}^{\underline{\mu}\nu\underline{\alpha_{1}\cdots\alpha_{l}}}= & 0,\text{ for any }i,j \,,
\end{align}
and $A$ is all terms which is antisymmetric in any two indices among
$\mu$ and $\alpha$'s. 
For identical reasons as the case for photons, only $A$ survive.
Thus, we can rewrite our amplitude in the $l$'th order as\footnote{Here we have used $q^{2}=0$ and drop out terms proportional to $q^{\mu}$,
which would not contribute to the gauge invariance amplitude.} 
\begin{align}
&M_{m+1,(l)}^{\mu\nu} \nonumber \\
= & \sum_{i=1}^{m} \frac{1}{(l+1)!}\frac{p_{i}^{\nu}}{p_{i}\cdot q}\left[p_{i}^{\mu}\left(q\cdot\partial_{i}\right)-\left(p_{i}\cdot q\right)\partial_{i}^{\mu}\right]\left(q\cdot\partial_{i}\right)^{l}M_{m}\nonumber \\
 & +q_{\alpha_{1}}\cdots q_{\alpha_{l}}A_{l}^{\underline{\mu}\nu\underline{\alpha_{1}\cdots\alpha_{l}}} \,.
\end{align}
For $l=0$, we get 
\begin{align}
M_{n+1,(0)}^{\mu\nu}=\sum_{i=1}^{m} \frac{p_{i}^{\nu}}{k\cdot q}q_{\alpha}J_{i}^{\mu\alpha}M_{m} \,,
\end{align}
but for $l>0$, we need to impose the gauge invariance condition again,
\begin{align}
\hspace{-0.1in} q_{\nu}\left(\sum_{i=1}^{m} \frac{1}{(l+1)!}\frac{p_{i}^{\nu}}{p_{i}\cdot q}\left[p_{i}^{\mu}\left(q\cdot\partial_{i}\right)-\left(p_{i}\cdot q\right)\partial_{i}^{\mu}\right]\right. \left(q\cdot\partial_{i}\right)^{l} & M_{m} \nonumber  \\
\left.+q_{\alpha_{1}}\cdots q_{\alpha_{l}}A_{l}^{\underline{\mu}\nu\underline{\alpha_{1}\cdots\alpha_{l}}}\right) & =0
\end{align}
Thus, we get 
\begin{align}
A_{l}^{\underline{\mu}\nu\underline{\alpha_{1}\cdots\alpha_{l}}} & =\nonumber \\
 &\hspace{-0.3in} -\sum_{i}\frac{1}{(l+1)!}\left(p_{i}^{\mu}\partial_{i}^{\alpha_{1}}-p_{i}^{\alpha_{1}}\partial_{i}^{\mu}\right)\partial_{i}^{\alpha_{2}}\cdots\partial_{i}^{\alpha_{l}}\partial_{i}^{\nu}M_{m}\nonumber \\
 &\hspace{-0.3in} +C_{l}^{\mu\nu\alpha_{1}\cdots\alpha_{l}}
\end{align}
For similar reasons, $C$ also contains only trace and antisymmetric part in any two
indices among $\nu$ and $\alpha$'s. Define $L^{\mu\nu}$ as
an antisymmetric tensor in $\mu$ and $\nu$, we can write 
\begin{align}
C_{l}^{\mu\nu\alpha_{1}\cdots\alpha_{l}}= & \sum_{i,j}L^{\mu\alpha_{i}}\left(L^{\nu\alpha_{j}}+\eta^{\nu\alpha_{j}}\right)D_{l}^{\alpha_{1}\cdots\tilde{\alpha_{i}}\cdots\tilde{\alpha_{j}}\cdots\alpha_{l}}\nonumber \\
 & +L^{\alpha_{i}\alpha_{j}}E_{l}^{\mu\nu\alpha_{1}\cdots\tilde{\alpha_{i}}\cdots\tilde{\alpha_{j}}\cdots\alpha_{l}},
\end{align}
so the amplitude becomes\footnote{The terms proportional to $q^{\nu}$ or $\eta^{\mu\nu}$ are dropped
out in gauge invariance amplitude.} \footnote{$\tilde{\alpha_{i}}$ means the entry $\alpha_{i}$ is removed.}
\begin{align}
M_{m+1,(l)}^{\mu\nu} & = \sum_{i=1}^{m} \frac{1}{(l+1)!}\frac{q_{\alpha}q_{\beta}}{p_{i}\cdot q}J_{i}^{\mu\alpha}J_{i}^{\nu\beta}\left(q\cdot\partial_{i}\right)^{l-1}M_{m}\nonumber \\
+ & q_{\alpha_{1}}\cdots q_{\alpha_{l}}\sum_{i,j}L^{\mu\alpha_{i}}L^{\nu\alpha_{j}}D_{l}^{\alpha_{1}\cdots\tilde{\alpha_{i}}\cdots\tilde{\alpha_{j}}\cdots\alpha_{l}},
\end{align}
For $l=1$, we do not have the $D$ terms, so we now get the sub-sub-leading
graviton soft theorem 
\begin{align}
M_{(1)}^{\mu\nu}= \sum_{i=1}^{m} \frac{q_{\alpha}q_{\beta}}{2k\cdot q}J_{i}^{\mu\alpha}J_{i}^{\nu\beta} \,.
\end{align}
However, for $l>1$, there are $D$ terms not given by Ward identity, 
\begin{align}
&\partial^{\alpha_{1}}\cdots\partial^{\alpha_{l}}M_{n+1,(l)}^{\mu\nu}\nonumber\\&\ \ =\partial^{\alpha_{1}}\cdots\partial^{\alpha_{l}}\nonumber\\&\ \ \ \ \ \ \left[\sum_{i=1}^{m}\frac{1}{(l+1)!}\frac{q_{\alpha}q_{\beta}}{p_{i}\cdot q}J_{i}^{\mu\alpha}J_{i}^{\nu\beta}\left(q\cdot\partial_{i}\right)^{l-1}\right]M_{n}\nonumber\\&\ \ \ \ +L^{\mu\alpha_{i}}L^{\nu\alpha_{j}}D_{l}^{\alpha_{1}\cdots\tilde{\alpha_{i}}\cdots\tilde{\alpha_{j}}\cdots\alpha_{l}}. 
\label{eq:1grav_inf}
\end{align}
Therefore, we can obtain, for example, either pieces symmetric in $\mu$ and $\alpha_i$, or those symmetric in $\nu$ and $\alpha_i$, 
\eqal{
 & \Omega_{\mu\alpha_{1}\cdots\alpha_{l}} \partial^{\alpha_{1}}\cdots\partial^{\alpha_{l}}M_{m+1,(l)}^{\mu\nu} \bigg|_{q\rightarrow 0} \nonumber \\
 & \ \ =\Omega_{\mu\alpha_{1}\cdots\alpha_{l}} \partial^{\alpha_{1}}\cdots\partial^{\alpha_{l}}\nonumber \\
 & \ \ \ \ \ \ \left[ \sum_{i=1}^{m} \frac{1}{(l+1)!}\frac{q_{\alpha}q_{\beta}}{p_{i}\cdot q}J_{i}^{\mu\alpha}J_{i}^{\nu\beta}\left(q\cdot\partial_{i}\right)^{l-1}\right]M_{m} \bigg|_{q\rightarrow 0}\nonumber \\
  & \Omega_{\nu\alpha_{1}\cdots\alpha_{l}} \partial^{\alpha_{1}}\cdots\partial^{\alpha_{l}}M_{m+1,(l)}^{\mu\nu} \bigg|_{q\rightarrow 0} \nonumber \\
 & \ \ =\Omega_{\nu\alpha_{1}\cdots\alpha_{l}} \partial^{\alpha_{1}}\cdots\partial^{\alpha_{l}}\nonumber \\
 & \ \ \ \ \ \ \left[ \sum_{i=1}^{m} \frac{1}{(l+1)!}\frac{q_{\alpha}q_{\beta}}{p_{i}\cdot q}J_{i}^{\mu\alpha}J_{i}^{\nu\beta}\left(q\cdot\partial_{i}\right)^{l-1}\right]M_{m} \bigg|_{q\rightarrow 0}.
}
where $\Omega^{\rho\alpha_{1}\cdots\alpha_{l}}$ is a totally symmetric tensor. 
The soft theorems from large gauge transformations \cite{gary}, however, only considers a more restrictive piece, 
\begin{align}
 & \left[ \Omega_{\mu \left( \nu\alpha_{1}\cdots\alpha_{l} \right)} + \Omega_{\nu \left( \mu\alpha_{1}\cdots\alpha_{l} \right)} \right] \partial^{\alpha_{1}}\cdots\partial^{\alpha_{l}}M_{m+1,(l)}^{\mu\nu} \bigg|_{q\rightarrow 0} \nonumber \\
 & \ \ =\left[ \Omega_{\mu \left( \nu\alpha_{1}\cdots\alpha_{l} \right)} + \Omega_{\nu \left( \mu\alpha_{1}\cdots\alpha_{l} \right)} \right]  \partial^{\alpha_{1}}\cdots\partial^{\alpha_{l}}\nonumber \\
 & \ \ \ \ \ \ \left[ \sum_{i=1}^{m} \frac{1}{(l+1)!}\frac{q_{\alpha}q_{\beta}}{p_{i}\cdot q}J_{i}^{\mu\alpha}J_{i}^{\nu\beta}\left(q\cdot\partial_{i}\right)^{l-1}\right]M_{m} \bigg|_{q\rightarrow 0},
\end{align}
where $\Omega_{\mu \left( \nu\alpha_{1}\cdots\alpha_{l} \right)}$ is totally symmetric in $\nu,\alpha_{1},\cdots,\alpha_{l}$, and traceless in all the indices \footnote{see footnote \ref{note:traceless} in the introduction.}. 
This follows from our result, but does not represent the most general derivable soft theorems. 

To consider an amplitude involving $n+1$ gravitons, 
\eqal{
M_{n+1} \left( q, k_1, \cdots, k_n \right)
}
we only have to replace the three-point vertex with the graviton self-interaction. 
The remaining steps are exactly the same. 
Taking one graviton soft, 
\eqal{
\hspace{-0.2in}M_{n+1}^{\mu\nu}=&\prod_{j=1}^{n}\epsilon_{j,\mu_{j}}\epsilon_{j,\nu_{j}}\sum_{i=1}^{n}\frac{V^{\mu\nu\mu_{i}\nu_{i}\alpha\beta}}{k_{i}\cdot q}M_{n,\alpha\beta}^{\mu_{1}\nu_{1}\cdots\tilde{\mu}_{i}\tilde{\nu}_{i}\cdots\mu_{n}\nu_{n}}\nonumber\\&+N^{\mu\nu},
\label{ngrav_polegut}
}
where $V$ is the graviton self-interaction vertex
\eqal{
V^{\mu\nu\mu_{i}\nu_{i}\alpha\beta}=&\left(k_{i}^{\mu}\eta^{\alpha\mu_{i}}+q_{\rho}\Sigma^{\rho\mu\alpha\mu_{i}}\right)\left(k_{i}^{\nu}\eta^{\beta\nu_{i}}+q_{\tau}\Sigma^{\tau\nu\beta\nu_{i}}\right) \nonumber \\
\Sigma^{abcd}=&\eta^{ac}\eta^{bd}-\eta^{ad}\eta^{bc}.
}
and $M_{n}$ is the amplitude involving the remaining $n$ gravitons. 
Again expand $N$ in $q$ and apply Ward identity, we can obtain, for $q^0$ order \footnote{The terms antisymmetric in $\mu$ and $\nu$ are dropped.}, the subleading soft theorem, 
\eq
M_{n+1,(0)}^{\mu\nu}=\sum_{i=1}^{n} {k_i^\nu \over k\cdot q} q_\alpha J_i^{'\mu \alpha} M_n \,,
\eqe
where
\eq
J_i^{'\mu\nu}=k_i^\mu \p_i^\nu - k_i^\nu \p_i^\mu + \epsilon_i^\mu {\p \over \p \epsilon_{i,\nu}} - \epsilon_i^\nu {\p \over \p \epsilon_{i,\mu}} \,.
\eqe
As for the $l$'th order where $l\ge 1$, the expansion of Eq.\,\eqref{eq:1grav_polegut} is 
\eqal{
&\hspace{-0.2in}M_{n+1,(l)}^{\mu\nu} \nonumber \\
=&\prod_{j=1}^{n}\epsilon_{j,\mu_{j}\nu_{j}} \sum_{i=1}^{n} \Bigg[\frac{k_{i}^{\mu}\eta^{\alpha\mu_{i}}k_{i}^{\nu}\eta^{\beta\nu_{i}}}{k_{i}\cdot q}\frac{\left(q\cdot\p_{i}\right)^{l+1}}{(l+1)!}\nonumber  \\
&+\frac{k_{i}^{\mu}\eta^{\alpha\mu_{i}}q_{\tau}\Sigma^{\tau\nu\beta\nu_{i}}+k_{i}^{\nu}\eta^{\beta\nu_{i}}q_{\rho}\Sigma^{\rho\mu\alpha\mu_{i}}}{k_{i}\cdot q}\frac{\left(q\cdot\p_{i}\right)^{l}}{l!}\nonumber\\
&+\frac{q_{\rho}\Sigma^{\rho\mu\alpha\mu_{i}}q_{\tau}\Sigma^{\tau\nu\beta\nu_{i}}}{k_{i}\cdot q}\frac{\left(q\cdot\p_{i}\right)^{l-1}}{(l-1)!}\Bigg]M_{n,\alpha\beta}^{\mu_{1}\nu_{1}\cdots\tilde{\mu}_{i}\tilde{\nu}_{i}\cdots\mu_{n}\nu_{n}}\nonumber \\
&+q_{\alpha_{1}}\cdots q_{\alpha_{l}}N_{l}^{\mu\nu\alpha_{1}\cdots\alpha_{l}} \,.
}
Applying Ward identity as before, the soft theorem is
\eqal{
&\hspace{-0.2in}M_{n+1,(l)}^{\mu\nu}\nonumber\\ 
=&\sum_{i=1}^{n}\frac{q_{\alpha}q_{\beta}}{k_{i}\cdot q}\left[\frac{J_{i}^{\mu\alpha}J_{i}^{\nu\beta}}{(l+1)!}+\frac{1}{2}\frac{J_{i}^{\mu\alpha}U_{i}^{\nu\beta}+U_{i}^{\mu\alpha}J_{i}^{\nu\beta}}{l!}\right.\nonumber\\ 
&\hspace{1.2in}\left.+\frac{1}{2}\frac{U_{i}^{\mu\alpha}U_{i}^{\nu\beta}}{(l-1)!}\right]\left(q\cdot\p_{i}\right)^{l-1}M_{n}\nonumber\\ 
&+q_{\alpha_{1}}\cdots q_{\alpha_{l}}\sum_{i,j}L^{\mu\alpha_{i}}L^{\nu\alpha_{j}}D_{l}^{\alpha_{1}\cdots\tilde{\alpha_{i}}\cdots\tilde{\alpha_{j}}\cdots\alpha_{l}} \,,
}
where
\eq
U_i^{\mu \nu} = \epsilon_i^\mu{\p \over \p \epsilon_{i,\nu}} - \epsilon_i^\nu{\p \over \p \epsilon_{i,\mu}} \,.
\eqe
In particular, the sub-sub-leading piece is
\eq
M_{n+1, (1)}^{\mu \nu} = \sum_{i=1}^{n} {q_\alpha q_\beta \over 2 k_i\cdot q} J_i^{'\mu\alpha} J_i^{'\nu\beta}
\eqe
without ambiguity. 
For $l>1$, we again have partially fixed soft theorem up to infinite order. 

%%%%%%%%%%%%%%%%%%%%%
\section{Example of Homogeneous Terms}
%%%%%%%%%%%%%%%%%%%%%

Here we show the anti-symmetric piece of $N_{l}^{\mu,\alpha_{1}\cdots\alpha_{l}}$
that was projected out is in fact non-zero, which means that the projected soft-theorem is indeed a ``partial soft theorem''.

\begin{figure}
\begin{subfigure}[b]{0.4\linewidth}
\includegraphics[width=\textwidth]{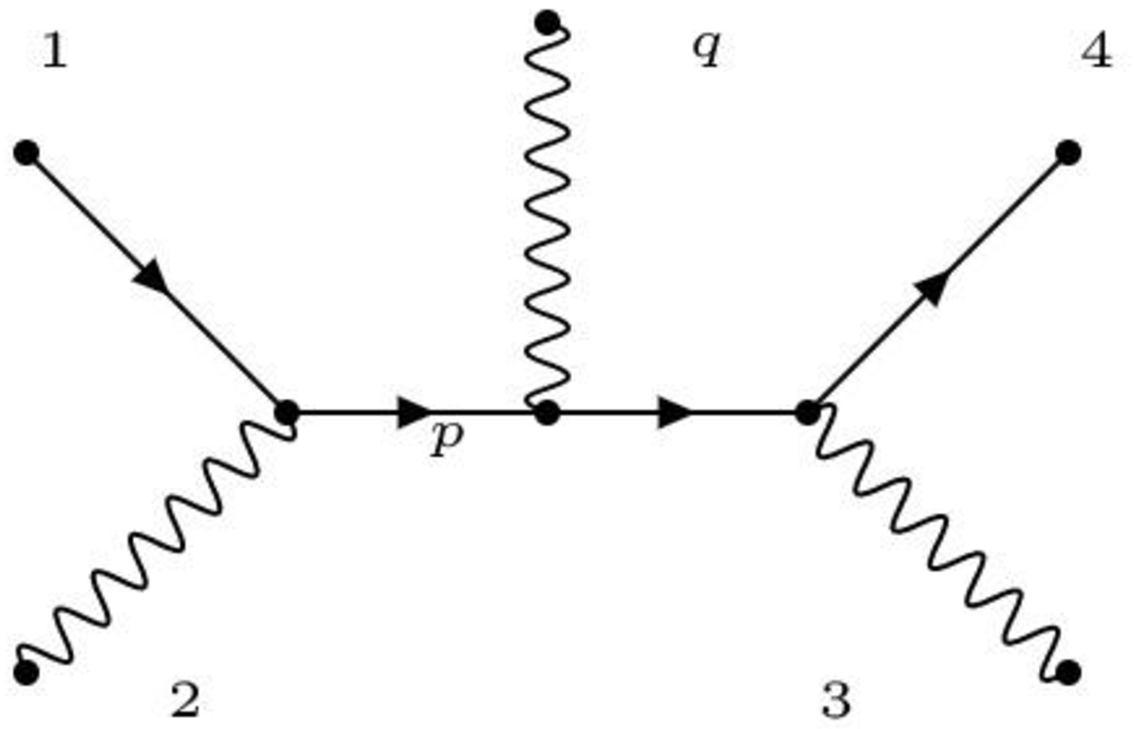}
\caption{} 
\label{figa}
\end{subfigure}
%%%
\begin{subfigure}[b]{0.4\linewidth}
\includegraphics[width=\textwidth]{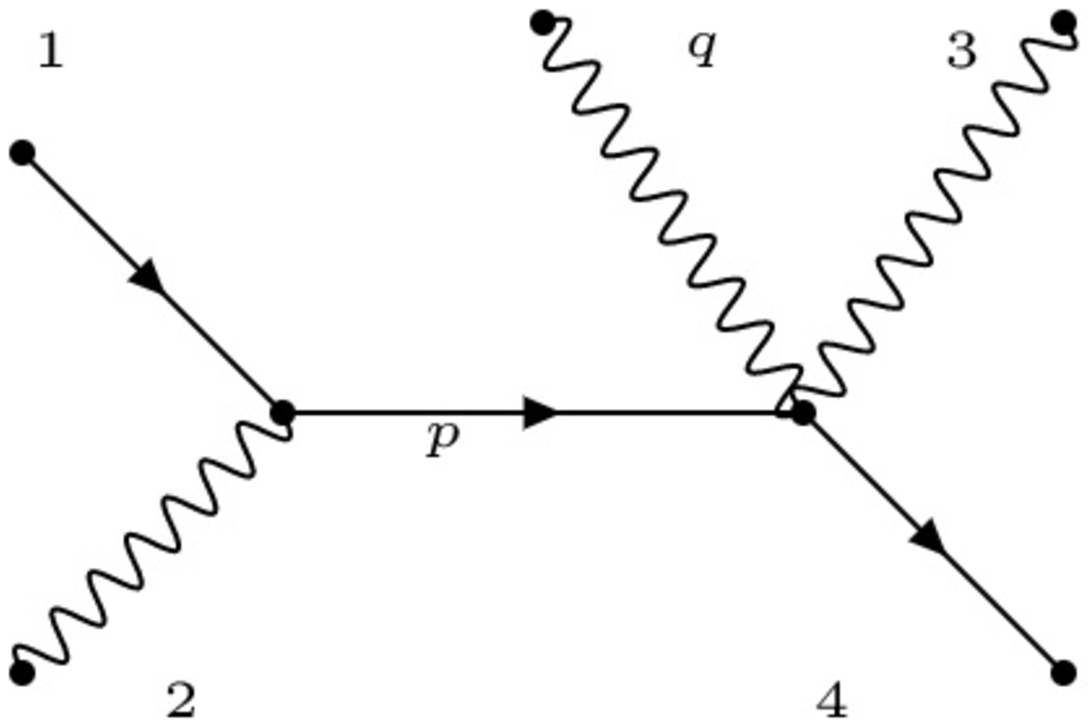}  
\caption{}
\label{figb}
\end{subfigure}
\caption{No pole diagrams of scalar QED five-point amplitude}
\label{fig:QED_5}
\end{figure}

We use an explicit scalar QED five-point amplitude to demonstrate.
The diagrams that contribute to $N$ comes from the soft photon coupled to an internal
leg as shown in the figure.
The contribution from Fig.~\ref{fig:QED_5}(a) is
\begin{align}
N&=(ie)\frac{(p_{1}+p)\cdot\epsilon_{2}}{(p_{1}+p)^{2}-m^{2}} \left[(ie)(p+p+q)\cdot\epsilon_{q}\right] \nonumber \\&(ie)\frac{(p+p_{4}+q)\cdot\epsilon_{3}}{(p+q)^{2}-m^{2}} +(2\leftrightarrow3,1\leftrightarrow4) \nonumber \\
 & =(ie)^{3}\frac{(p_{1}+p)\cdot\epsilon_{2}}{(p_{1}+p)^{2}-m^{2}}(2p\cdot\epsilon_{q})\frac{(p+p_{4}+q)\cdot\epsilon_{3}}{(p+q)^{2}-m^{2}} \nonumber \\&+(2\leftrightarrow3,1\leftrightarrow4) =\epsilon_{q,\mu}N^{\mu}
\end{align}where $p=p_{1}+p_{2}$, and $(2\leftrightarrow3,1\leftrightarrow4)$ means we have to sum
over (2,3) and (1,4) exchange.

Then we perform derivative on $N^{\mu}$, then anti-symmetrize the $(\mu,\alpha_{1})$
index
\begin{align}
N^{\mu, \alpha_{1}}&=
\frac{\partial}{\partial q_{\alpha_{1}}}  N^{\mu} \nonumber \\&=(-ie)^{3}\frac{(p_{1}+p)\cdot\epsilon_{2}}{(p_{1}+p)^{2}-m^{2}}\frac{\partial}{\partial q^{\alpha_{1}}}\left[(2p)^{\mu}\frac{(p+p_{4}+q)\cdot\epsilon_{3}}{(p+q)^{2}-m^{2}}\right] \nonumber \\
 & =(-ie)^{3}\frac{(p_{1}+p)\cdot\epsilon_{2}}{(p_{1}+p)^{2}-m^{2}} [2\frac{p^{\mu}\epsilon_{3}^{\nu}}{(p+q)^{2}-m^{2}} \nonumber \\& + 2\frac{p^{\mu}p^{\alpha_{1}}(p+q)\cdot\epsilon_{3}}{\left[(p+q)^{2}-m^{2}\right]^{2}}]=S^{\mu\alpha_{1}}+A^{\mu \alpha_{1}}
\end{align} The anti-symmetric part $A^{\mu \alpha_{1}}$ is not zero.

The contribution from Fig.~\ref{fig:QED_5}(b) is 
\begin{align}
N & =(ie)\frac{(p_{1}+p)\cdot\epsilon_{2}}{(p_{1}+p)^{2}-m^{2}}(-2ie^{2})(\epsilon_{2}\cdot\epsilon_{q})+(q\leftrightarrow3,1\leftrightarrow4)
\end{align}However, this term doesn't contribute to the $N^{\mu, \alpha_{1}}_{1}$ since it doesn't
involve $q$. After considering the Fig.~\ref{fig:QED_5}(a) and Fig.~\ref{fig:QED_5}(b), we have shown the anti-symmetric part of $N^{\mu, \alpha_{1}}_{1}$ is non-zero, but at the end we drop this term to obtain the partial soft theorem. 

%%%%%%%%%%%%%%%%%%%%%%%%
\section{Effect of Higher Dimensional Operators}
%%%%%%%%%%%%%%%%%%%%%%%%

Now we consider the soft photon theorem in the effective field theory \cite{elvang}, where the sub-leading ($q^{0}$) soft photon theorem will be modified
in the presence of the effective operator. 
The effective operator starts to contribute at $q^{0}$ order and continues to affect higher order ones, so we will explicitly show its modification to the infinite order soft theorem.

Here is the modification for sub-leading soft photon theorem,
\begin{align}
& M_{n+m+1}|_{q\rightarrow 0}=(q^{-1}\mathcal{S}^{(-1)}+q^{0}\mathcal{S}^{(0)})M_{n+m}\nonumber \\&+q^{0}\widetilde{\mathcal{S}}^{(0)}\widetilde{M}_{n+m}+\mathcal{O}(q^{1})
\end{align}where the tilde on the n-point amplitudes indicates that the particle
type of the $k$th leg of $\widetilde{M}_{n+m}$ may differ from that
in $\widetilde{M}_{n+m+1}$.

\begin{figure}[h]
\includegraphics[height=1.6cm]{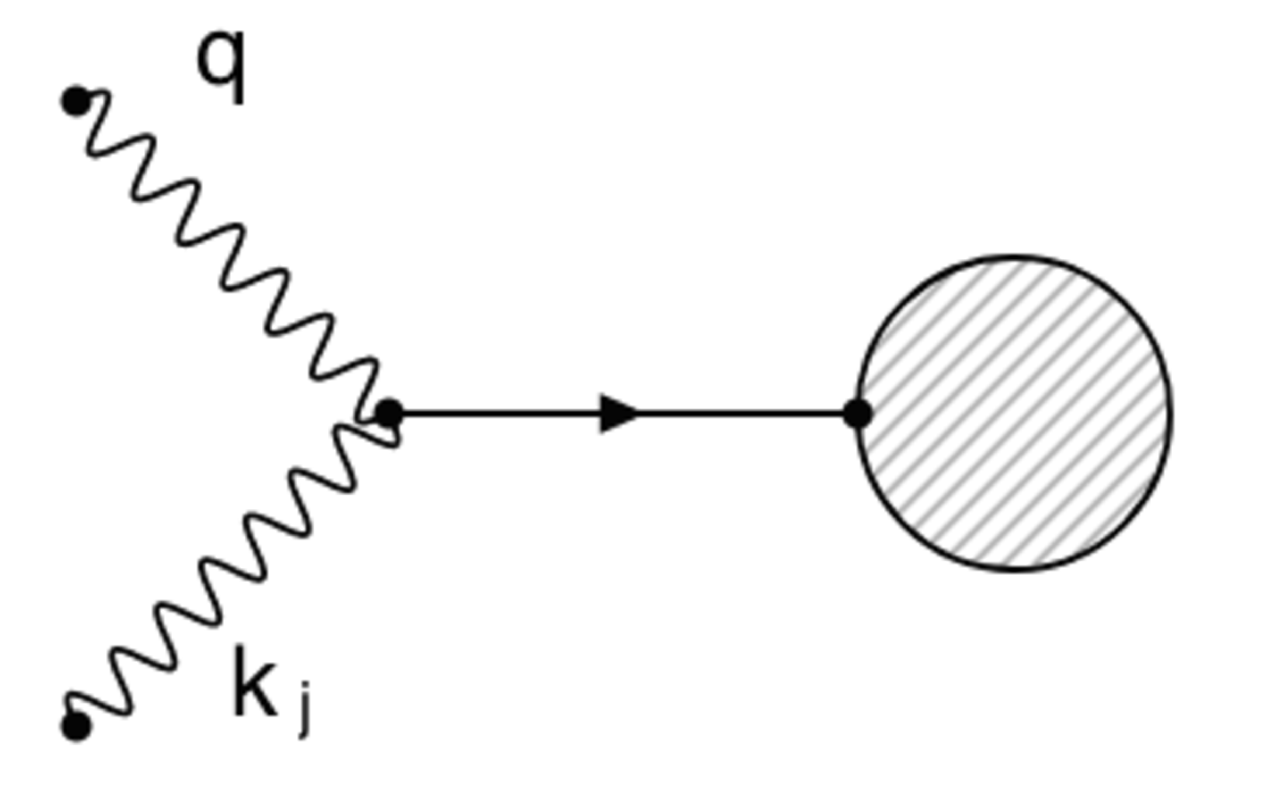}
\caption{} 
\label{fig_qk}
\end{figure}

We choose a specific effective operator, $\varphi F^{\mu\nu}F_{\mu\nu}$ ($\varphi$ is a real scalar field),
to show its explicit form of modification. When one of external leg is taken soft, the internal $\varphi$
propagator goes on-shell and the amplitude factorizes as shown in the Fig.~\ref{fig_qk}.

\noindent Its contribution is
\begin{align}
g\left[2(k_{j} \cdot q)(\epsilon_{k_{j}}\cdot\epsilon_{q})-2(k_{j} \cdot\epsilon_{q})(q\cdot\epsilon_{k_{j}})\right]\frac{1}{2(k_{j}\cdot q)}\widetilde{M}_{n+m}
\end{align}with g the coupling constant for the three-point vertex.

We have shown the effective operator starts making contribution at sub-leading ($q^{0}$) order, and we now discuss how it modifies the infinite soft theorem.
Again we separate the pole diagrams and the no pole ones, then see how Ward identity gives constraints at each $q$ order.
\begin{align}
0 & =q_{\mu}M_{n+m+1}^{\mu}|_{q\rightarrow 0}\nonumber \\
 & =\sum_{i=1}^{m} e_{i}M_{n+m}\nonumber \\&+\sum_{j=1}^{n}g\left[(k_{j}\cdot q)(\epsilon_{k_{j}}\cdot q)-(k_{j}\cdot q)(q\cdot\epsilon_{k_{j}})\right]\frac{1}{(k_{j}\cdot q)}\widetilde{M}_{n+m} \nonumber \\
 &+q_{\mu} N^{\mu} \label{eq:three_term} \\
 & =\sum_{i=1}^{m} e_{i}M_{n+m}+q_{\mu}N^{\mu} \label{eq:two_term}
\end{align}The first term in Eq.\,\eqref{eq:three_term} is the original pole diagram from photon and matter field coupling, the second one is the pole diagram from effective operator, and the third one are no pole diagrams.
We find that the pole diagram from effective operator doesn't constrain the no pole diagram since itself is gauge invariant.
So Eq.\,\eqref{eq:two_term} is basically the same as Eq.\,\eqref{eq:Ward_N}.
The effective operators doesn't constrain the form of $N$, but it still modifies the infinite soft
theorem to be
\begin{align}
 & \Omega_{\mu\alpha_{1}\cdots\alpha_{l}} \partial^{\alpha_{1}}\cdots\partial^{\alpha_{l}}M_{n+m+1}^{\mu}|_{q\rightarrow 0}\nonumber \\
 & = \Omega_{\mu\alpha_{1}\cdots\alpha_{l}} \Bigg\{ \sum_{i=1}^{m}\frac{1}{(l+1)!}\frac{e_{i}}{p_{i}\cdot q}q_{\nu}J_{i}^{\mu\nu}\partial_{i}^{\alpha_{1}}\cdots\partial_{i}^{\alpha_{l}}M_{n+m} \nonumber \\
 &  +\sum^{n}_{j=1}g\left[(k_{j}\cdot q)\epsilon_{k_{j}}^{\mu}-(q\cdot\epsilon_{k_{j}})k_{j}^{\mu} \right]\frac{1}{(k_{j}\cdot q)} \partial_{i}^{\alpha_{1}}\cdots\partial_{i}^{\alpha_{l}} \widetilde{M}_{n+m}] \Bigg\}
\end{align}

%%%%%%%%%%%%%%%%%%%%%%%%
\section{Conclusion and Discussion}
%%%%%%%%%%%%%%%%%%%%%%%%

In this letter, we demonstrate how on-shell gauge invariance can fix higher order soft limit of photons and gravitons up to an undetermined homogeneous term $R^{\mu}$. 
This leads to infinite order soft theorems on certain projected pieces of amplitude, to which the homogeneous term does not contribute. 
We explicitly worked out the appropriate projection to obtain such pieces, and showed that the infinite order soft theorems derived from large gauge transformations can be completely reproduced here. 
For the case of gravitons, the theorems derived here are actually more complete, leaving fewer undetermined pieces in the amplitude. 

We use explicit examples to demonstrate that the homogeneous term in $R^{\mu}$ can be projected out but can be non-zero, which means we indeed drop some to obtain the infinite order soft theorem.
Finally, we consider the effect of adding higher dimensional operator, which starts to modify photon soft theorem at sub-leading order.
Moreover, its modification to the infinite order soft theorem can also be obtained.

The fact that the soft-theorems derived from residual gauge symmetries, so far  can all be reproduced by ordinary on-shell gauge symmetry, leaves us asking what is the relevance of this new symmetry on a physical observable like the S-matrix. A pessimist may say that the evidence so far is that there are no relevance beyond that implied by ordinary gauge symmetry, which in a sense is not surprising given that one projects the correlation function to obtain the S-matrix and thus certain information might be projected out. Alternatively, one might say that the symmetry is in fact telling us that we are using the wrong asymptotic states for the S-matrix and thus ignorant to its features. 

We choose single particle states for the S-matrix due to it being irreducible representations of the Poincare group. This statement makes no distinction between massless and massive kinematics. However, for massless kinematics, it is well known that single particle states are ill-defined, as there are no quantum numbers available for us to differentiate colinear multi-particle states, and manifest itself in the IR divergence of massless scattering amplitudes. Thus perhaps the infinite residual gauge symmetry is telling us that the correct asymptotic state for massless kinematics should form representation of this infinte group. Indeed recent analysis along this line for QED has demonstrated that this indeed appears to be the case~\cite{Kapec}, albeit a similar analysis for gravity is still lacking. It will be interesting to understand this in full generality and illustrate how modifications of the three-point interaction via higher dimension operators changes the conclusion. 

Besides single soft theorems discussed here, 
one may apply the method in \cite{us} to consider double soft theorems, which involve two, instead of one, soft gauge bosons. 
It would be interesting to see whether such theorems can be similarly pushed to infinite order by considering a projected piece of the amplitude. 

%%%%%%%%%%%%%%%%%%%%%%%%
\section{Acknowledgement}
%%%%%%%%%%%%%%%%%%%%%%%%
We thank Yu-tin Huang for suggesting the problem and helping with the draft. Zhi-Zhong Li, Hung-Hwa Lin and Shun-Qing Zhang are supported by MoST grant 106-2628-M-002-012-MY3.

%%%%%%%%%%%%%

\end{document}